# Investigation of Long Monolayer Graphene Ribbons grown on Graphite Capped 6H-SiC (000-1)

Nicolas Camara,[1] Jean-Roch Huntzinger,[2] Gemma Rius,[1] Antoine Tiberj,[2] Narcis Mestres,[3] Francesc Pérez-Murano,[1] Philippe Godignon[1] and Jean Camassel[2]

[1] IMB-CNM-CSIC, Campus UAB 08193-Bellaterra, Barcelona, Spain

[2] GES, UMR-CNRS 5650, Université Montpellier 2, 34095-Montpellier cedex 5, France

[3] ICMAB-CSIC, Campus UAB 08193-Bellaterra, Barcelona, Spain

nicolas.camara@cnm.es, jean-roch.huntzinger@univ-montp2.fr, Gemma.Rius@cnm.es, antoine.tiberj@ges.univ-montp2.fr, Francesc.Perez@cnm.es, narcis.mestres@icmab.es, Philippe.Godignon@cnm.es, camassel@ges.univ-montp2.fr

ABSTRACT: We present an investigation of large, isolated, graphene ribbons grown on the C-face of on-axis semi-insulating 6H-SiC wafers. Using a graphite cap to cover the SiC sample, we modify the desorption of the Si species during the Si sublimation process. This results in a better control of the growth kinetics, yielding very long (about 300 μm long, 5 μm wide), homogeneous monolayer graphene ribbons. These ribbons fully occupy unusually large terraces on the step bunched SiC surface, as shown by AFM, optical microscopy and SEM. Raman spectrometry indicates that the thermal stress has been partially relaxed by wrinkles formation, visible in AFM images. In addition, we show that despite the low optical absorption of graphene, optical differential transmission can be successfully used to prove the monolayer character of the ribbons.

MANUSCRIPT TEXT: Graphene has emerged recently as a new material with outstanding electronic properties. This includes mass-less Dirac fermions, ballistic transport properties at room temperature and good compatibility with the silicon planar technology. Graphene-based devices are then promising candidates to complement silicon in the future generations of high frequency microelectronic devices. Different techniques have been developed over the past 4 years to fabricate mono or bi-layers of graphene. They range from exfoliated graphite, either mechanically,[1] or in a liquid-phase solution,[2] to chemical vapor deposition on a metal surface,[3,4] and more recently, to substrate-free synthesis when passing ethanol into an argon plasma.[5] The method investigated in this work consists in a controlled sublimation of few atomic layers of Si from a mono crystalline SiC substrate.[6] Such epitaxial growth (EG) of graphene seems to be the most suitable option for industrial applications,[7] but for easy control, it necessitates large and homogeneous sheets of monolayer or few layers of graphene (FLG) covering either a full-wafer surface or a specific area, like an open window in an AlN pre-patterned SiC substrate.[8]

Unfortunately, whatever the SiC polytype under investigation (4H, 6H or even 3C), the sublimation conditions pressure varying from UHV (Ultra-High Vacuum) below $10^{-9}$ Torr to more standard SV (Secondary Vacuum) conditions in the range of $10^{-8}$ to $10^{-6}$ Torr, it is still challenging to grow FLG with homogeneous domain sizes larger than few hundred nanometers.[9-11] However, sublimation from the C-face leads to wider domains and higher mobility than the Si-face.[10] But still, it is hardly possible to process homogeneous devices on a wafer. Noticeable exceptions are the recent results by Virojanadara *et. al.*[12] and Emtsev *et al.*[13] that showed that performing graphitization on a 6H-SiC substrate under Argon at 900 mbar could lead to large, regular, graphene monolayers and bilayers, but on the Si-face. The purpose of this work is twofold. First, we show that there is an alternative way in which very large monolayer graphene ribbons can be grown on the C-face of a graphite-capped SiC sample using only a commercial radio frequency (RF) heated furnace under SV conditions. Secondly, we show that the combination of Raman spectroscopy coupled with differential transmission constitutes a most powerful tool to investigate the thickness homogeneity of the growth product.

All samples were 1x1 cm$^2$ templates cut from a 3-inch, on-axis, semi-insulating 6H-SiC wafer from Cree Research. Before cutting, electrochemical polishing was done by Novasic to get Epiready® morphology.[14] A sacrificial oxide was then thermally grown, and chemically etched in HF to remove any (small) sub-surface damage from the polishing process. The necessary chemical treatments were clean-room compatible and very similar to the ones used before thermal oxidation or post-implantation annealing in standard SiC technology. Atomically flat surfaces were obtained in this way. The vacuum limit in the RF furnace was 10$^{-6}$ Torr and, before sublimation, the samples were heated at 1150°C for 10 min in order to remove any trace of native oxide. With the aim to investigate the influence of the graphite cap, two different series of samples (A and B) have been grown.

Samples A were grown at 1550°C for 5 minutes under SV. We know already that such growth conditions allow the formation of FLG flakes over the whole SiC sample but not homogeneously.[7,15,16] This is because the process is not just intrinsic (simple sublimation of Si and reorganization of the C atoms to form graphene). In most cases, defects facilitate graphitization. Recently, it has been suggested that threading dislocations are efficient nucleation centers.[15] Using OM (Optical Microscopy) in the crossed polarization mode and in dark field mode, we evidence mainly two different growth features (see Figure 1a and 1b).

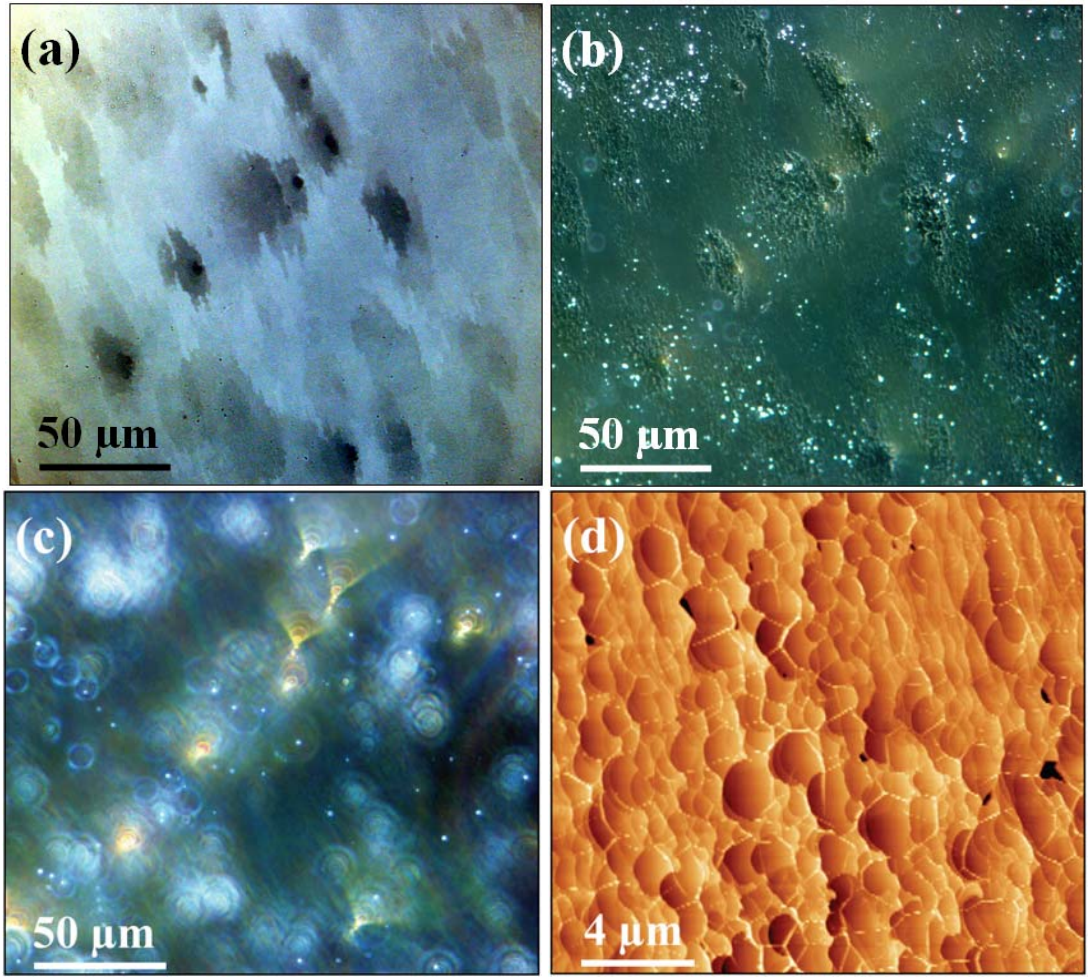

**Figure 1.** (a) Wide range, crossed-polarization optical microscopy of samples A, fully covered with FLG. Growth conditions: 1550°C under SV, without graphite cap. Thicker FLG areas appear darker. (b) Same area, dark field optical microscopy, showing the yellow dislocation spots that originate the thicker FLG areas. (c) Same as (b), except a slight defocusing allowing a better localization of the dislocations. (d) Typical AFM image of samples A taken in the grey sea region of the picture (a).

First is the fast growth of tens of graphene layers nucleated around the pre-existing defects, like dislocations. Those dislocations are visible in dark field mode as yellow spots in the center of the thick flakes (Figure 1b). They are even more visible by defocusing the microscope up to few microns under the surface where they appear as yellow cones (see Figure 1c).[15] Next, surrounding these darker/thicker areas are thinner FLGs which form the grey sea. The mechanism leading to these FLGs was the second one discussed in ref [15], and will be called the "grey sea process" in the following. These FLG form the well known structure shown by AFM in Figure 1d with an average domain size which hardly reaches 1 µm wide.[7,17,18] Such a growth process is not expected to produce a large and homogeneous pavement of graphene over the whole SiC wafer. That is why a new, radically different growth process has been developed. The aim was to quench the grey sea growth process rather than to optimize it, and to rely on the control of the growth assisted by defects.

This was done on samples B by increasing the C and Si partial pressure near the SiC surface, covering the sample with a graphite cap. It resulted, at 1550°C, in the quenching of the whole graphitic growth process. Raising the temperature up to 1700°C, we found out that the grey sea growth was still quenched while very long graphene ribbons up to 300 µm long and 5 µm wide, surrounded by SiC, sparsely covered the surface. One of these ribbons is shown in Figure 2a by OM in dark field mode. At these conditions, the growth process is totally different.
First, the dislocations visible as yellow cones in Figure 2b are not as efficient nucleation sites as before, otherwise all of them would be a starting point for a ribbon. Moreover, we found many ribbons with no dislocation nearby. On the other hand, defects were often found at the center of the ribbon. When probed by Raman spectroscopy, we found out that these defects have a very intense typical signature of disordered graphitic material. More work is needed to fully characterize the origin of this extrinsic nucleation.
Second, the SiC surface reorganizes in much longer terraces. This step-bunching, standard in SiC technology is related to a small initial miscut of the wafer surface with respect to the nominal 6H-SiC surface. It seems that graphene is allowed to expand quickly along these terraces, which explains the significant length of the ribbons.
Third, the initial SiC steps edges, in the range of 2 nm high, stop the graphene growth perpendicular to the terraces. As a result, the graphene layer expands preferentially only on one terrace as shown by AFM on Figures 2b and 2c. The terrace edges coincide with the edges of the ribbon. The graphene-covered terrace is systematically much wider than the surrounding ones. The most probable is that the graphene layer stabilizes the terrace, allowing it to expand. This expansion occurs at the expense of the surrounding terraces.

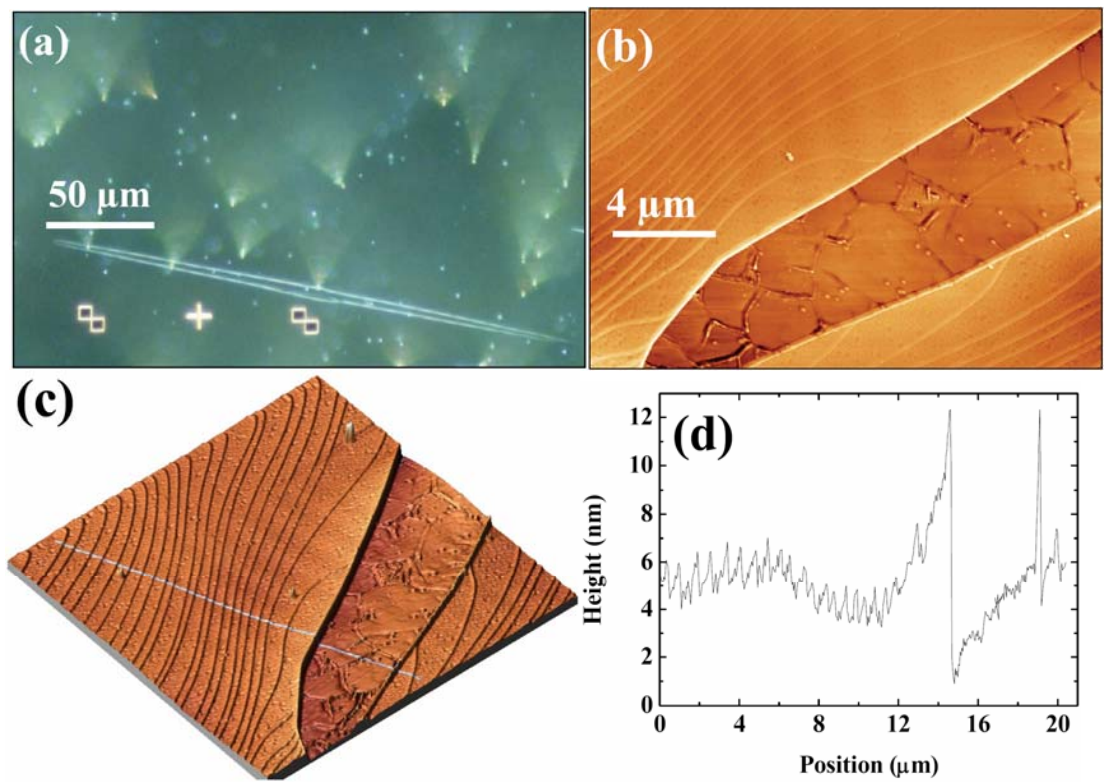

**Figure 2.** (a) Wide range Optical Microscopy in dark field view of samples B annealed at 1700°C under SV covered by a graphite cap. The C-face surface of this 6H-SiC on axis sample is covered by very long graphene ribbons (b) AFM picture in phase mode of a graphene monolayer surrounded by the typical SiC step bunching.(c) AFM topography 3D picture of the same area (d) and it's profile showing clearly the SiC step bunching between 1 and 2 nm high and the very high step between the SiC surface and the terrace where lay the graphene layer.

At the end of the process, the ribbon is surrounded by sharp and high edges (more than 10nm), as indicated by AFM in Figure 2c and 2d. Contrary to graphene exfoliated on an oxidized Si wafer, epitaxial graphene grown on SiC is almost invisible by OM.[19] What is actually seen in Figure 2a or in Figure 3a is not the graphene itself, but these sharp edges surrounding the ribbon. The SEM image on Figure 3b confirms, on a larger scale, the conclusions of the AFM investigations discussed above: the edges of the graphene ribbon coincide with the edges of the terrace. The graphene layers are flat and exhibit only wrinkles as usually found on graphitic material grown on SiC.[17] As it can be seen on the AFM pictures in Figure 2b, these wrinkles are few nanometers high and come from the weak interaction between the graphene ribbon and the underlying buffer layer. Large atomically flat and ripple-free areas in the range of several µm² are systematically detected.

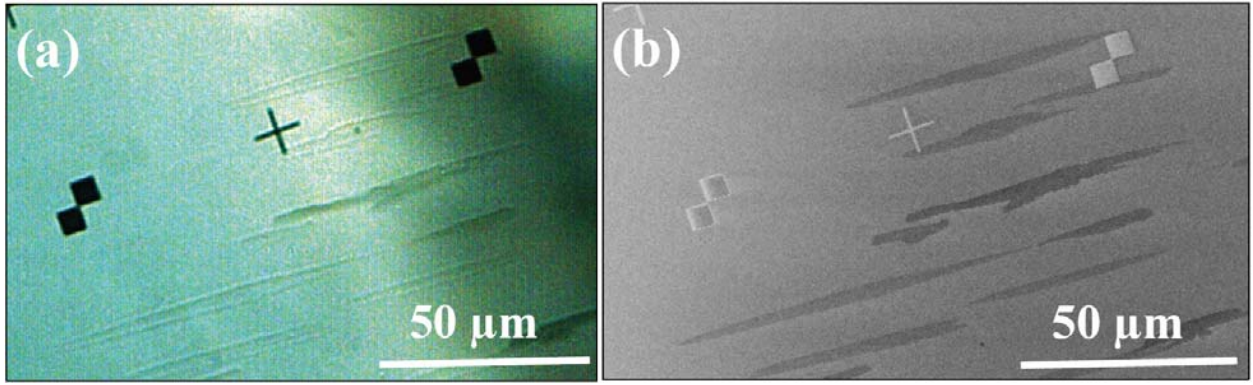

**Figure 3.** (a) Optical Microscopy image in Nomarski crossed polarization mode of the large graphene islands grown at 1700°C on a sample covered by graphite cap. The graphene monolayers are transparent, hardly visible. (b) SEM image of the same area. The graphene monolayers are light-grey while the multilayers are darker. The crosses and squares are deposited gold marks.

The quality and uniformity of these ribbons were investigated by Raman spectroscopy and differential transmission experiments. Raman spectra were collected at room temperature, in confocal configuration using the 514 nm line of an Ar⁺-ion laser as excitation. As usual, the bare SiC reference has been subtracted. The ribbons are homogeneous with the typical signature shown on Figure 4c. These spectra are very close to the ones found in the literature for graphene monolayer exfoliated on top of $SiO_2$/Si.[20-26] Indeed, the 2D-band has a symmetrical band shape that can be fitted by a single Lorentzian with a FWHM (Full-Width at Half Maximum) ranging from 28 to 34 cm⁻¹. It falls at low frequency between 2690 and 2705 cm⁻¹. Furthermore, the G-band falls at high frequency between 1595 and 1605 cm⁻¹, with a FWHM that ranges from 10 to 15 cm⁻¹. The ratio of integrated intensities $I_{2D}/I_G$ ranges from 3 to 5. If it is assumed that graphene layers behave as those exfoliated on $SiO_2$, these high upshift of the G-band positions would require a high level of doping,[23] but then the G-band FWHM should be less than 9 cm$^{-1}$. So, there is a residual compressive strain, though much smaller than observed in other works.[27-29] For our high growth temperatures, a high differential dilation stress would be expected instead. This stress has surely been relaxed by the formation of the wrinkles seen by AFM in figure 2. Only in few cases, a weak D-band appears at low frequency (around 1350 cm$^{-1}$), probably coming from the wrinkles, charged impurities or original defect. These observations are fully compatible with high quality, slightly strained, monolayer graphene. Nevertheless, they only prove that the ribbons are not multilayers with AB (Bernal) stacking. It is not enough to conclude that they are monolayer. Indeed, it has been shown that Raman measurements may result in identical spectral shapes for monolayer graphene and misoriented multilayers graphene on $SiO_2$.[30,31]

In this work, we propose a simple and very reliable technique that allows to discriminate between both cases. It consists in coupling differential transmission measurement with Raman spectroscopy. We measured with a high quality power meter the laser power transmitted through the sample, under the microscope, possibly during the Raman spectrum acquisition. This is possible since SiC is transparent to the laser, contrary to oxidized silicon substrates. The transmittance of a graphene layer on top of a substrate has already been calculated.[32] The point is to consider a single interface between two media (for instance air/substrate), with boundary conditions for the electromagnetic field that are modified by the presence of the conducting graphene layer. In our case, the electrical permittivities of the two media are $\varepsilon_1 = 1.0$ and $\varepsilon_2 = 7.20$ for air and SiC substrate respectively. The optical conductivity of a graphene monolayer is $\sigma = e^2/4\hbar$.[32] This prediction has been confirmed from transmission measurement through suspended monolayer graphene.[33] Using the eq. (50) in ref. 32, these values lead to T = 0.7814. Without graphene (bare SiC substrate), the transmittance is $T_0 = 0.7912$. The corresponding relative extinction $\eta = (T_0-T)/T_0$ is then 1.23 % for a graphene monolayer. A bilayer has an optical conductivity twice as large as a monolayer in the visible range[34] leading to a relative extinction of 2.44 %. Let us note that actual transmittance of the sample depends also on the back side SiC/air interface, which is optically polished in our case. Nevertheless, this simply involves a common factor that cancels out, so that η depends only on the relative change of the transmittance through the first interface.

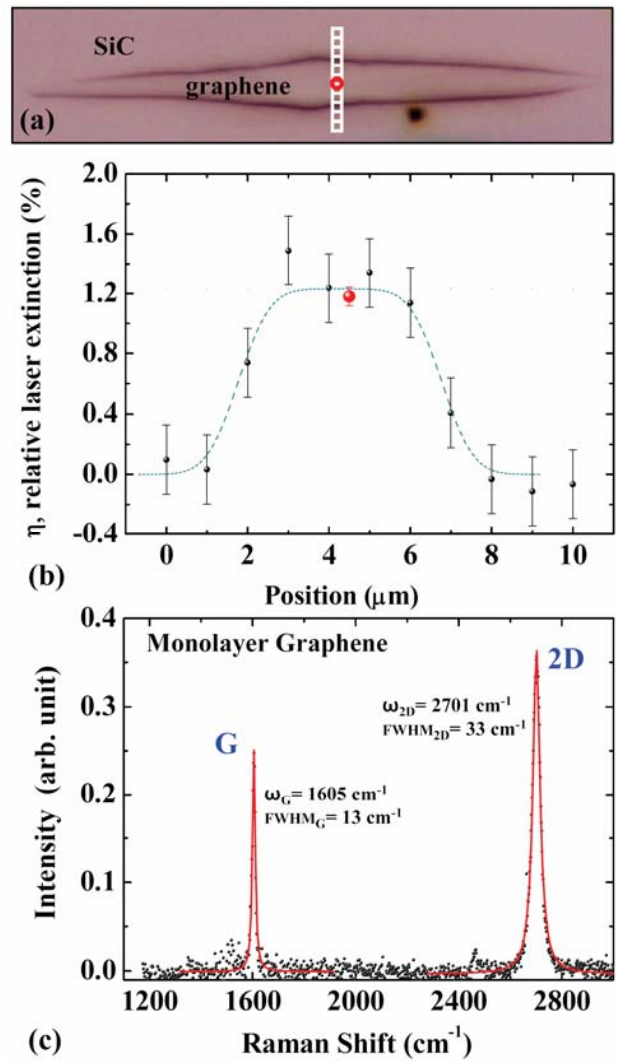

**Figure 4.** (a) Graphene ribbon from sample B visible by OM in dark field mode used for the coupled Raman/differential transmission measurements. (b) Black point: linear scan of $\eta = (T_0-T)/T_0$ along the 10 μm white bar shown in (a); dash line: the theoretical value for a graphene monolayer on SiC calculated as the convolution of the 4μm wide ribbon by a Gaussian laser beam (FWHM = 1 μm); red point: acquired simultaneously with the Raman spectrum shown in (c), at the position marked as a red circle in (a) at the center of the ribbon.

As an example of our techniques, we selected one graphene ribbon and we performed differential transmission measurements along the 11 different points shown in Figure 4a. The relative extinction η is plotted versus the laser spot position in Figure 4b. The shape of the plot confirms that the spot size is small enough to accurately measure the extinction of the ribbon. The noise related to the power measurement has been evaluated from successive acquisitions. Several differential transmission measurements have been taken during the Raman spectra acquisition at the center position of the graphene ribbon (red circle in Figure 3a) and on the bare SiC substrate. The longer time acquisition lead to an increase of the accuracy, giving η = 1.18 ± 0.06 % (coverage factor of 3). Without any ambiguity, this confirms that the layer shown in Figure 4 is a true monolayer graphene ribbon.

To summarize, we have shown that using a graphite cap to lower the desorption of Si species, and working at high temperature (~1700°C), very long graphene ribbons can be grown on the C-face of a SiC wafer. These ribbons are up to 300 μm long and 5 μm wide, all oriented in the same direction and with wrinkles-free domains of several μm$^2$. It was demonstrated by combining AFM, optical microscopy and SEM that these ribbons fully occupy a single terrace of the step-bunched SiC surface. Raman spectroscopy indicates high quality, slightly strained, homogeneous ribbons. In addition, we have shown that optical differential transmission can be successfully used to prove the monolayer character of the ribbons. We expect this technique to spread widely as a companion tool for Raman, when working on transparent substrates as SiC.

ACKNOWLEDGMENT: We greatly acknowledge the EC for partial support through the RTN "ManSiC" Project. One of us (N.C.) also acknowledges the Spanish Government for a Grant "Juan de la Cierva 2006". Finally, partial support of the Spanish Consolider NANOSELECT (CSD2007-00041) is also acknowledged.